\documentclass{iau} 
\usepackage{graphicx}
\usepackage{euscript}
\usepackage[normalem]{ulem}
\usepackage{natbib}
\usepackage{hyperref} 
\hypersetup{
    colorlinks,
    linkcolor={blue},
    citecolor={blue},
    urlcolor={blue}
}

\newcommand{\mocca}{{\sc mocca}}
\newcommand{\Msun}{M_{\odot}}

\def\apj{{\it ApJ}~} 
\def\apjl{{\it ApJL}~} 
\def\mnras{{\it MNRAS}~} 
\title[] 
{Are most Cataclysmic Variables
in Globular Clusters 
dynamically formed?}
\author[D. Belloni et al.]   
{
Diogo Belloni$^{1,2}$,
Mirek Giersz$^3$,
Liliana E. Rivera Sandoval$^4$,
Abbas Askar$^5$
\and
Pawel Ciecielag$^3$ 
}

\affiliation{
$^1$National Institute for Space Research,
S\~ao Jos\'e dos Campos, Brazil\\ 
[\affilskip]
$^2$Instituto de F{\'i}sica y Astronom{\'i}a, 
Universidad de Valpara{\'i}so, 
Valpara{\'i}so, Chile \\
email: {\tt diogo.belloni@inpe.br}\\
[\affilskip]
$^3$Nicolaus Copernicus Astronomical Centre,
Polish Academy of Sciences,
Warsaw, Poland\\
email: {\tt mig@camk.edu.pl}\\
[\affilskip]
$^4$Department of Physics and Astronomy, 
Texas Tech University,
Lubbock, USA\\
email: {\tt Liliana.Rivera@ttu.edu}\\
[\affilskip]
$^{5}$Lund Observatory, 
Lund University, Lund, Sweden\\
email: {\tt askar@astro.lu.se}
}

\pubyear{2019}
\volume{351}  
\jname{Star Clusters: From the Milky Way to the Early Universe}
\editors{A. Bragaglia, M.B. Davies, A. Sills \& E. Vesperini, eds.}
\begin{document}

\maketitle

\begin{abstract}
We have been investigating populations of 
cataclysmic variables (CVs) in a set of 
more than 300 globular cluster (GC) models 
evolved with the \mocca\,code.
One of the main questions we have intended 
to answer is whether \textit{most CVs in GCs
are dynamically formed or not}.
Contrary to what has been argued for a 
long time, we found that dynamical 
destruction of primordial CV progenitors 
is much stronger in GCs than dynamical 
formation of CVs.
In particular, we found that, on average, 
the detectable CV population is predominantly 
composed of CVs formed via a typical 
common envelope phase
($\gtrsim$ 70 per cent).
However, core-collapsed models tend to 
have higher fractions of bright CVs than 
non-core-collapsed ones, which suggests 
then that the formation of CVs is indeed 
slightly favoured through strong dynamical 
interactions in core-collapsed GCs, due to 
the high stellar densities in their cores.
\end{abstract}

\firstsection 

\section{Motivation and Approach}

Cataclysmic variables (CVs) 
are interacting binaries harbouring
a white dwarf (WD) undergoing
dynamically and thermally stable 
mass transfer from a low-mass 
companion, usually a main-sequence 
(MS) star 
\citep[e.g.][]{Warner_1995_OK}.
They are expected to exist in 
non-negligible numbers in 
globular clusters (GCs), 
which are natural laboratories 
for testing theories of stellar 
dynamics and evolution. 

Due to the high stellar crowding 
in GCs and and the CV intrinsic 
faintness, CVs are difficult 
to identify in such environments.
%
%
Until now the best-studied GCs 
with respect to CV populations are 
NGC 6397 \citep{Cohn_2010}, 
NGC 6752 \citep{Lugger_2017}, 
$\omega$ Cen \citep{Cool_2013,Henleywillis_2018} 
and
47 Tuc \citep[e.g.][]{Rivera_2017}.
The identification of CVs in these 
GCs has been carried out by 
identifying the 
\textit{Hubble Space Telescope} 
(\textit{HST})
optical counterparts to 
\textit{Chandra} X-ray sources.
Usually these counterparts show an 
H$\alpha$ excess (suggesting the 
presence of an accretion disc), 
they are bluer 
than the MS stars and several 
also show photometric variability 
in different bands.

In the core-collapsed
clusters NGC 6397 and NGC 6752, 
CVs can be divided into 
two populations, a bright and 
a faint one.
On their optical colour-magnitude 
diagrams (CMDs), {\it bright} CVs 
lie close to the MS and
{\it faint} CVs close to (or on) 
the WD cooling sequence,
R $\approx$ 21.5 mag being the 
cut-off between both populations.
Interestingly, in the non-core-collapsed 
clusters 47 Tuc and $\omega$ Cen,
only one CV population is observed,
that being mainly composed of faint CVs.
In addition, the observed number of bright 
CVs per cluster mass in core-collapsed 
clusters is so far much higher than in non-core-collapsed clusters.

In the series of papers by 
\citet{Belloni_2016a,
Belloni_2017a,
Belloni_2017b,
Belloni_2019}, 
we have analysed a large set of 
$\sim300$ GC models with 
a focus on the properties of
their present-day CV populations, 
CV progenitors,  and how CV 
properties are affected 
by dynamics in dense environments. 
The prime goal in this series of
papers is to explain the observed
properties and to answer the 
following questions:
i) {\it Are most CVs in GCs
dynamically formed? }
ii) {\it Do we need to invoke dynamics
to explain main properties of bright 
and faint CVs?}
We summarize here the main findings
in this series with focus on the 
influence of dynamics on CV progenitor
destruction and CV formation. 
%
%

In order to simulate the GC models, 
we used the MOnte Carlo Cluster simulAtor 
({\sc mocca}) code developed by 
\citet{Giersz_2013} and
\citet[][and references therein]{Hypki_2013},
which includes the {\sc fewbody} code 
\citep{Fregeau_2004} to perform
numerical integrations of three- or 
four-body gravitational interactions and
the Binary Stellar Evolution ({\sc bse})
code \citep{Hurley_2002}, 
with the upgrades described in 
\citet{Belloni_2018b}
to deal with CV evolution.
%
%
%
Amongst our models, we have a variety 
of different initial conditions spanning 
different values of the 
mass, 
size, 
King concentration parameter, 
initial binary population
\citep[e.g.][]{Belloni_2017c,Belloni_2018a},
binary fraction, 
and 
Galactocentric distances.
In addition, we also explored two parameters
of stellar/binary evolution, namely inclusion 
or not of mass fallback for BH formation
\citep{Belczynski_2002}
and three different common
envelope phase (CEP) efficiencies.
All variables in our modelling 
(i.e. binary evolution parameters, 
initial binary populations, 
and 
initial cluster conditions) 
are summarized in table 1 in 
\citet{Belloni_2019}.


\vspace{-0.25cm}

\section{Dynamical Destruction versus
Dynamical Formation}

\begin{figure*}
   \begin{center}
    \includegraphics[width=0.48\linewidth]{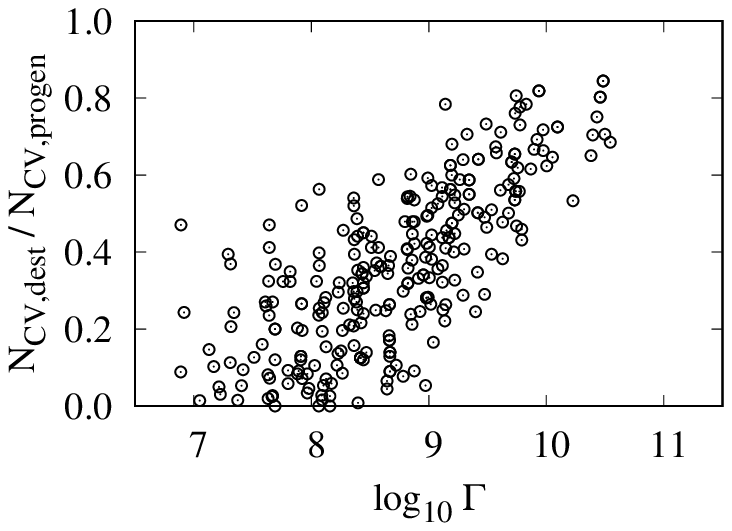}
    \includegraphics[width=0.48\linewidth]{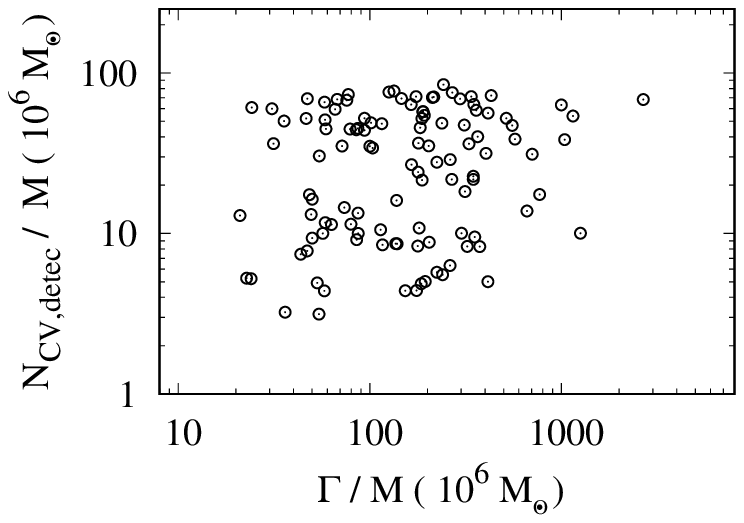}
    \end{center}
  \caption{Fraction of dynamical destroyed CV
   progenitors
   ($N_{\rm CV,dest}/N_{\rm CV,progen}$) against
   the initial stellar encounter rate
   $\Gamma$ (left-hand panel)
   and mass-normalized number of 
   detectable CVs (right-hand panel) 
   against the present-day 
   mass-normalized $\Gamma$.
   Note that there is a clear correlation
   between the fraction of destroyed CV
   progenitors and the initial $\Gamma$.
   On the other hand, 
   there is no (or very weak, if at all) 
   statistical evidence for a 
   correlation between the detectable
   CV abundance and the present-day 
   $\Gamma$.
   }
  \label{Fig}
\end{figure*}

We start discussing the rate of 
destruction of primordial CV 
progenitors with respect to the 
initial stellar encounter rate
given by 
$\Gamma = \rho_0^2 R_c^3 \sigma_0^{-1}$
\citep[e.g.][]{Pooley_2006,Hong_2017},
where $\rho_0$, $R_c$ and $\sigma_0$ 
are the central density, the core radius 
and the mass-weighted central velocity 
dispersion, respectively.
%
%
We quantify the fraction of primordial 
CV progenitors that are destroyed in 
dynamical interactions before becoming 
CVs ($N_{\rm CV,dest}/N_{\rm CV,progen}$).
This is illustrated in 
the left-hand panel of Fig.~\ref{Fig}, 
where we show such a fraction versus 
$\Gamma$ for all models. 
Note that the greater the $\Gamma$, 
the higher the fraction of destroyed 
primordial CV progenitors (i.e. the 
stronger the influence of dynamical 
interactions on destroying these 
progenitors).
In terms of the soft-hard boundary, the 
greater the $\Gamma$,  the shorter the 
period that defines the boundary between 
soft and hard binaries.
We carried out Pearson's rank correlation 
tests, and we found strong correlation 
with more than 99.9 per cent confidence,
being $r\approx0.78$.

Regarding the formation 
channels, \citet{Belloni_2019} 
found that the dominant one amongst 
detectable CVs is typical CEP  
($\approx88^{+12}_{-18}$ per cent, 
for both core-collapsed and 
non-core-collapsed clusters).
We also found that the average 
fraction of dynamically formed 
CVs among only bright CVs is 
relatively low ($\approx9_{-9}^{+24}$ 
per cent, for both core-collapsed and 
non-core-collapsed clusters).
In other words, we found here no 
(or very weak, if at all) correlation 
between the number of either detectable 
CVs or bright CVs with respect to the 
cluster type (e.g. related to the 
stellar encounter rate).

Our findings are in agreement 
with recent studies of 
\textit{Chandra} X-ray sources in 
GCs by \citet{Cheng_2018}. 
Using a sample of 69 GCs and
focusing on CVs and chromospherically
active binaries,
these authors found that there is not 
a significant correlation between 
the number of X-ray sources 
and the mass-normalized stellar 
encounter rate. 
These findings disagree with previous 
results, which considered smaller 
GC samples \citep[e.g.][]{Pooley_2006}. 
A correlation would be expected 
if dynamical interactions largely 
influence the creation of X-ray 
sources.
However, \citet{Cheng_2018} have 
shown that dynamical interactions 
are less dominant than previously 
believed, and that the primordial 
formation has a substantial 
contribution. 
In the right-hand panel of 
Fig. \ref{Fig} we show the number 
of detectable CVs normalized by 
the total cluster mass in units 
of $10^6~\Msun$ as a function 
of $\Gamma$ also normalized by 
the total cluster mass in units 
of $10^6~\Msun$. 
We found no (or very weak, if at all) 
statistical evidence for a correlation 
between the mass-normalized CV abundance 
and the mass-normalized $\Gamma$.

The physical reason for this is 
associated with the role of dynamics
in creating and destroying pre-CVs.
We notice that destruction of CV 
progenitors takes place mainly for 
MS-MS binaries during the first few 
hundred Myr of cluster evolution. 
Later, when WD-MS binaries are created, 
dynamical interactions are very 
strongly suppressed, because during 
the CEP there is a substantial 
reduction of binary periods.
Regarding dynamical pre-CV formation, 
there are three main possible 
scenarios:
i) {\it interaction between a low-mass 
MS-MS binary and a single WD},
ii) {\it interaction between a low-mass 
MS-MS binary and a single MS}, and
iii) {\it interaction between a 
WD-MS binary and a single MS}.
We found that the main scenarios 
previously proposed in the literature
for dynamical formation of faint and bright
CVs in GCs have a very low probability of 
occurring, which explains our findings with 
respect to the influence of dynamics on CV 
formation (very low fraction of dynamically 
formed faint and bright CVs) and 
with respect to the stellar encounter 
rate (no/extremely weak correlation 
with the amount of detectable and 
bright CVs).

With respect to the fraction of 
bright CVs among detectable ones,
we found here that, on 
average,  for non-core-collapsed 
models, $\sim 5-30$ per cent of 
the detectable CVs are bright,
which is consistent with 47 Tuc 
and $\omega$ Cen.
Regarding core-collapsed models, 
we found that fraction to be 
$\sim5-45$ per cent.
However, our core-collapsed
clusters with the shortest 
half-mass relaxation times usually 
have bright CV fractions higher than 
$\sim50$ per cent.
This is consistent with observational 
results of NGC 6397 and 
NGC 6752, which have fractions of 
bright CVs in the range of 
$\sim40-60$ per cent.
Our results suggest then that the 
formation of CVs is indeed slightly 
favoured through strong dynamical 
interactions in core-collapsed GCs,
especially those with very short 
half-mass relaxation times.

\vspace{-0.25cm}

\section{Conclusions}

We have analysed a relatively large 
sample  of more than 300 GC models, 
evolved with an up-to-date 
version of the \mocca\,code, with respect to 
initial binary populations and 
stellar/binary evolution prescriptions.
%
%
%
We found a strong correlation at a significant 
level between the fraction of destroyed
primordial CV progenitors and the initial 
stellar encounter rate, i.e. we found
that the greater the initial stellar encounter 
rate, the stronger the role of dynamical 
interactions in destroying primordial CV 
progenitors.
In addition, we showed that dynamical destruction 
of primordial CV progenitors is much stronger 
in GCs than dynamical formation of CVs.
Moreover, we found that the detectable CV 
population is predominantly composed
of CVs formed via typical CEP 
($\gtrsim 70$ per cent).
Finally, even though amongst detectable 
CVs the fractions of bright/faint CVs 
change from model to model, we found that, 
on average, non-core-collapsed models 
tend to have small fractions of bright CVs, 
while core-collapsed models have higher 
fractions, so that dynamics play a 
sufficiently important role to explain 
the relatively larger fraction of 
bright CVs observed in core-collapsed 
clusters.
%

\vspace{0.25cm}

%
\textbf{Acknowledgements}:
DB was supported by the grants \#2017/14289-3, \#2013/26258-4 and \#2018/23562-8, S\~ao Paulo Research Foundation (FAPESP).
MG acknowledges partial support from the National Science Center, Poland, through the grant UMO-2016/23/B/ST9/02732.
AA is supported by the Carl Tryggers Foundation for Scientific Research through the grant CTS 17:113.


\end{document}